\documentclass[a4paper,11pt]{article}
\usepackage{pos}
\usepackage{subfigure}

\title{Fire streaks, electromagnetic effects, directed flow and lifetime
of the plasma at SPS energies}

\author*[a]{Vitalii Ozvenchuk}
\author[a,b]{Antoni Szczurek}
\author[a]{Andrzej Rybicki}

\affiliation[a]{H.~Niewodnicza\'nski Institute of Nuclear Physics, Polish Academy of Science,\\
 Radzikowskiego 152, 31-342 Krak\'ow, Poland}

\affiliation[b]{University of Rzesz\'ow,\\
 Rejtana 16c, 35-959 Rzeszów, Poland}
 
\emailAdd{Antoni.Szczurek@ifj.edu.pl}
\emailAdd{Andrzej.Rybicki@cern.ch}
\emailAdd{Vitalii.Ozvenchuk@ifj.edu.pl}

\abstract{We present our calculation of electromagnetic effects~\cite{EM_paper}, induced by the spectator charge on Feynman-$x_F$ distributions of charged pions in peripheral $Pb+Pb$ collisions at CERN SPS energies, including realistic initial space-time-momentum conditions for pion emission. The calculation is performed in the framework of a specific implementation of the fire-streak model, adopted to the production of both $\pi^-$ and $\pi^+$ mesons. Isospin effects are included to take into account the asymmetry in production of $\pi^+$ and $\pi^-$ at high rapidity. A comparison to a simplified model from the literature is made. We obtain a good description of the NA49 data on the $x_F$- and $p_T$-dependence of the ratio of cross sections $\pi^+/\pi^-$. The experimental data favors short times ($0.5<\tau<2$~fm/$c$) for fast pion creation in the local fire-streak rest frame. The possibility of the expansion of the spectators is considered in our calculation, and its influence on the electromagnetic effect observed for the $\pi^+/\pi^-$ ratio is discussed. The influence of directed and elliptic flow, and vorticity on the observed effect is also estimated. We conclude that the fire-streak model, which properly describes the centrality dependence of $\pi^-$ rapidity spectra at CERN SPS energies, also provides realistic initial conditions for pion production. Consequently, it provides a quantitative description of the electromagnetic effect on the $\pi^+/\pi^-$ ratio as a function of $x_F$.}

\FullConference{%
  40th International Conference on High Energy physics - ICHEP2020\\
  July 28 - August 6, 2020\\
  Prague, Czech Republic (virtual meeting)
}

\begin{document}
\maketitle

\section{Introduction}
We study the electromagnetic effects in heavy-ion collisions, in particular, in peripheral $Pb+Pb$ collisions at CERN Super Proton Synchrotron (SPS) energies. After the collision the positively charged spectators generate electromagnetic fields, which modify the trajectories of charged pions. We use this effect as a new source of the information on the space-time evolution of the system.

The NA49 experimental data on Feynman-$x_F$ dependence of $\pi^+/\pi^-$ ratio~\cite{NA49_EM} shows a strong electromagnetic effect especially for the pions with small transverse momentum in the vicinity of spectator rapidity. These experimental results can be qualitatively reproduced within the Monte Carlo model by introducing the emission distance $d_E$~\cite{EM_previous_1}. It defines the initial distance between the pion emission region, which is reduced in this model to a single point in space, and the spectator. A reasonably optimal description of the data was obtained when the original pion source was not far from the spectator. The best, even if still qualitative, agreement with the experimental data~\cite{NA49_EM} was obtained for $d_E\approx0.5-1$~fm.

In this contribution, largely based on our recent paper~\cite{EM_paper}, we address the question whether models with more realistic initial conditions can describe the electromagnetic effects observed for $\pi^+$ and $\pi^-$ spectra. It was shown that one can describe the broadening of the $\pi^-$ rapidity distribution with centrality, as observed by the NA49 experiment~\cite{NA49_rapidity}, within a special implementation of the fire-streak model which we proposed in Ref.~\cite{FS_model_1}. In our opinion, this specific version of the fire-streak model provides realistic initial conditions for quark-gluon plasma creation at SPS energies, in particular obeying energy-momentum conservation.

\section{The model}
\begin{figure}
\centering
\includegraphics[width=.6\textwidth]{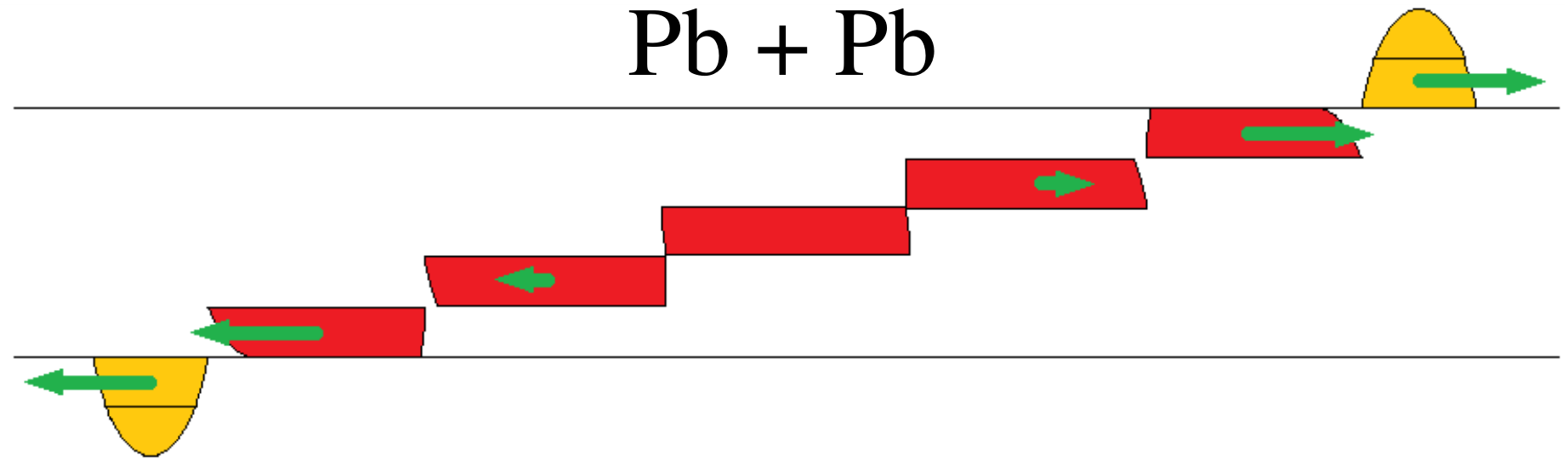}
\caption{The situation after the collision. The area marked in red shows the partonic matter. Each element moves with a different longitudinal velocity, which can be obtained from energy-momentum conservation~\cite{FS_model_1}.}
\label{schematic_collision}
\end{figure}
The details about our version of fire-streak model were given in Refs.~\cite{FS_model_1,FS_model_2}. We will mention here only the main concepts of the model. The model describes the longitudinal evolution of the system. We divide the colliding nuclei in transverse plane into two-dimensial 1~fm $\times$ 1fm bricks. After the collision the fire streaks are formed (see Fig.~\ref{schematic_collision}). The fire streaks are moving along the collision axis. Each fire streak fragments independently into pions by the single fire-streak fragmentation function parametrized in Ref.~\cite{FS_model_1} in the form:
\begin{equation}
\frac{dn}{dy}(y,y_s,E_s^*,m_s)=A(E_s^*-m_s)\exp\biggl(-\frac{[(y-y_s)^2+\epsilon^2]^{\frac{r}{2}}}{r\sigma_y^r}\biggr),
\label{fragmentation_function}
\end{equation}
where $y$ is the rapidity of the pion, $y_s$ is the fire-streak rapidity given by energy-momentum conservation, $E_s^*$ is its total energy in its own rest frame, and $m_s$ is the sum of ``cold'' rest masses of the two nuclear ``bricks'' forming the fire streak (see Ref.~\cite{FS_model_1}). The free parameters of the function~(\ref{fragmentation_function}) are $A$, $\sigma_y$ and $r$, which are fitted to NA49 experimental data~\cite{NA49_rapidity} for the rapidity distribution of $\pi^-$. It appeared that these parameters are common to all fire streaks and independent of $Pb+Pb$ collision centrality. Our model describes the whole centrality dependence of rapidity distributions of $\pi^-$ including, in particular, the narrowing of the rapidity distribution from peripheral to central $Pb+Pb$ collisions.

Our model provides only the information about the longitudinal expansion of the system, thus for the initial transverse-momentum distribution of pions we choose that obtained from UrQMD model~\cite{UrQMD}, which well describes the NA49 data~\cite{NA49_rapidity} on transverse-momentum mass spectrum of $\pi^-$ produced at midrapidity in peripheral $Pb+Pb$ collisions at top SPS energy. The resulting UrQMD predictions for transverse-momentum distributions of pions we then parametrize by the exponential function:
\begin{equation}
\label{fit_function}
\frac{dN}{dp_T}=\frac{Sp_T}{T^2+mT}\exp{[-(m_T-m)/T]},
\end{equation}
where $m$ is the mass of the pion, $m_T=\sqrt{m^2+p_T^2}$ is its transverse mass, $S$ and $T$ are the yield integral and the inverse slope parameter, respectively. The fit to the UrQMD simulations at midrapidity gives $T_{\pi^-}=165$~MeV and $T_{\pi^+}=163$~MeV. 

\begin{figure*}
\vspace{0.3cm} \centering \subfigure{
\resizebox{0.47\textwidth}{!}{%
 \includegraphics{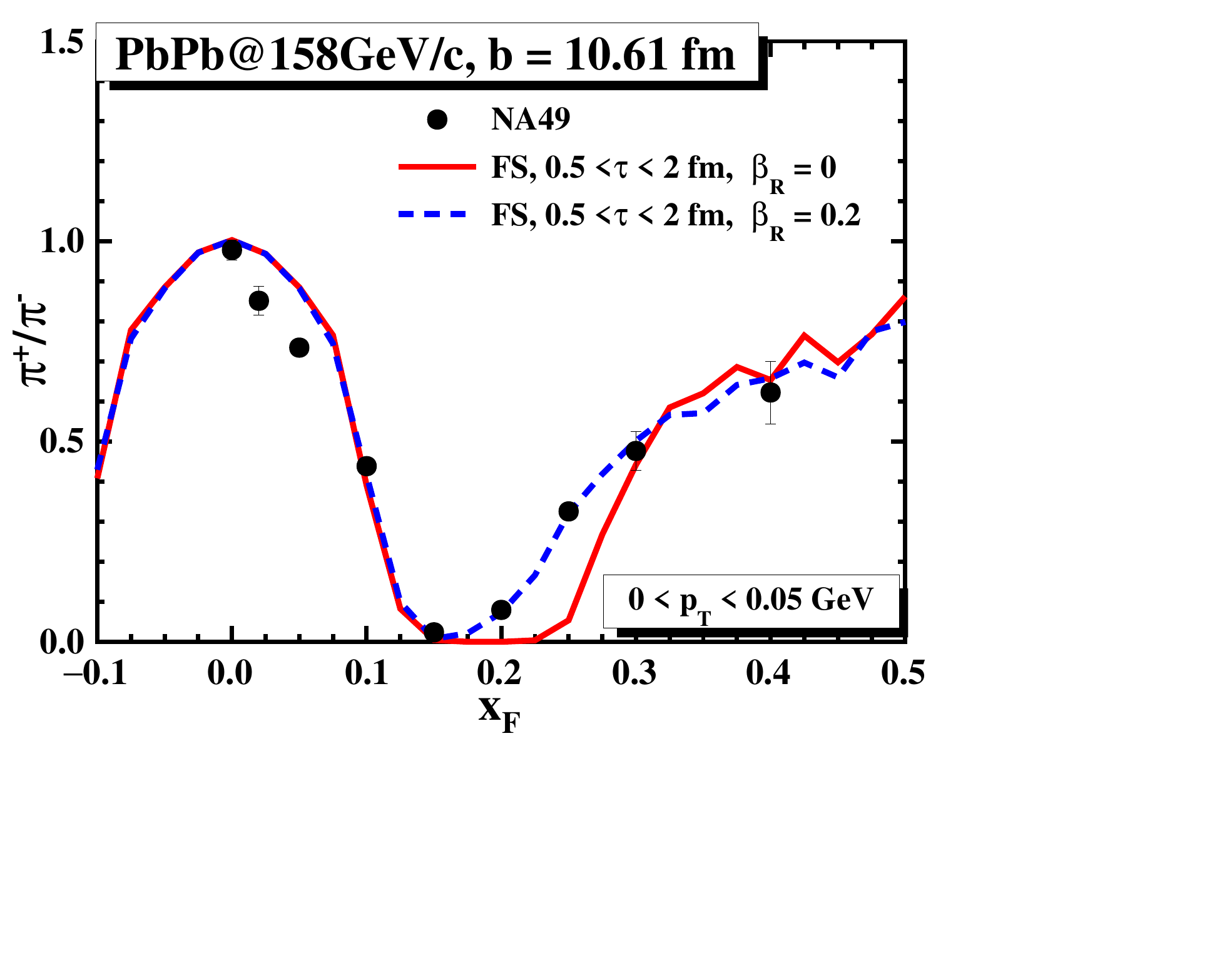}\hspace*{-5cm}
} } \subfigure{
\resizebox{0.47\textwidth}{!}{%
 \includegraphics{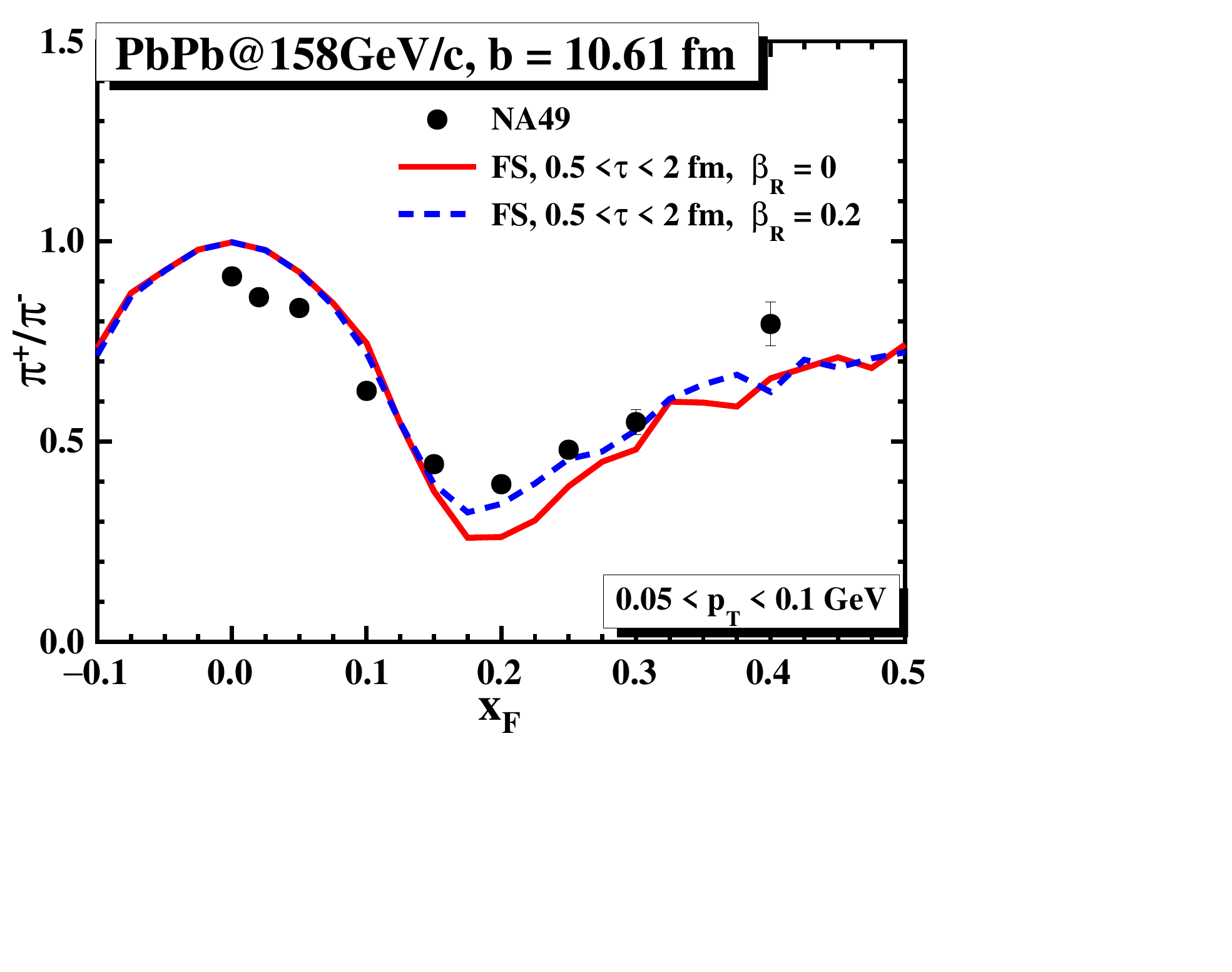}\hspace*{-5cm}
} } \vspace{-1.7cm}\caption{The results of calculation of
electromagnetic effects on the $\pi^+/\pi^-$ ratio as a function of
$x_F$ in peripheral $Pb+Pb$ collisions at top SPS energy, obtained
within our version of the fire-streak model with the pion
creation time given by Eq.~(\ref{tau_dependence}) for two different
ranges of pion transverse momentum: $0<p_T<0.05$~GeV/$c$ (left plot)
and $0.05<p_T<0.1$~GeV/$c$ (right plot). The scenario with stable
spectators is shown by the solid red lines, and the scenario with
expanding spectators with the radial surface velocity,
$\beta_R=0.2$, is presented by the dashed blue lines.} \label{final}
\end{figure*}

In the present contribution we quantify the possible effect of pion flow on the observed $\pi^+ / \pi^-$ ratio. The NA49 collaboration clearly observed~\cite{NA49_flow} nonzero elliptic ($v_2$) and directed ($v_1$) pion flow coefficients. We parametrize experimental data on $v_2$ and $v_1$ as a function of pion rapidity and transverse momentum. For our electromagnetic effect we need data in a broad range of $x_F$, i.e. at rather large rapidities. Therefore we included also the data of the WA98 collaboration~\cite{WA98_flow} into the fit.

To start the calculation of electromagnetic effects we have to fix also the time of emission of pions from the fire streaks and the initial position of the pion relative to the spectators. This cannot be calculated from first principles and it is treated here as a free parameter. We introduce the pion emission time, $\tau$, in the fire-streak rest frame. Up to this time the fire streak evolves in the longitudinal direction. For our calculations of electromagnetic effect we have to calculate the actual position of pion creation in $z$ for each ($i,j$) fire streak in the nucleus-nucleus center-of-mass system. This can be done by applying the Lorentz transformation. The transformation is given by the velocity of a given fire streak
in the center-of-mass system which in turn is given by its position in the impact parameter plane $(b_x,b_y)$~\cite{FS_model_1}.

\begin{figure*}
\vspace{0.3cm} \centering \subfigure{
\resizebox{0.47\textwidth}{!}{%
 \includegraphics{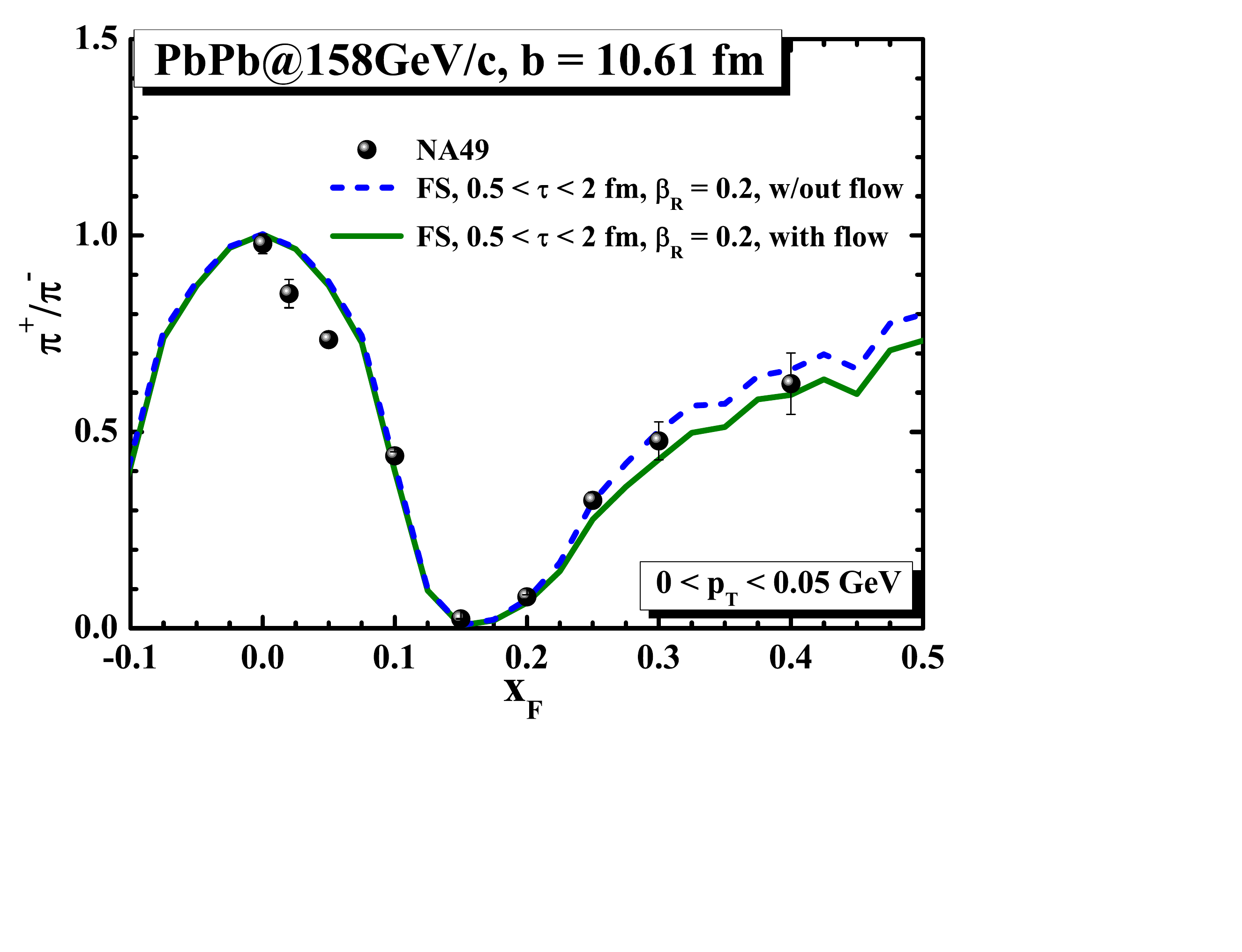}\hspace*{-6cm}
} } \subfigure{
\resizebox{0.47\textwidth}{!}{%
 \includegraphics{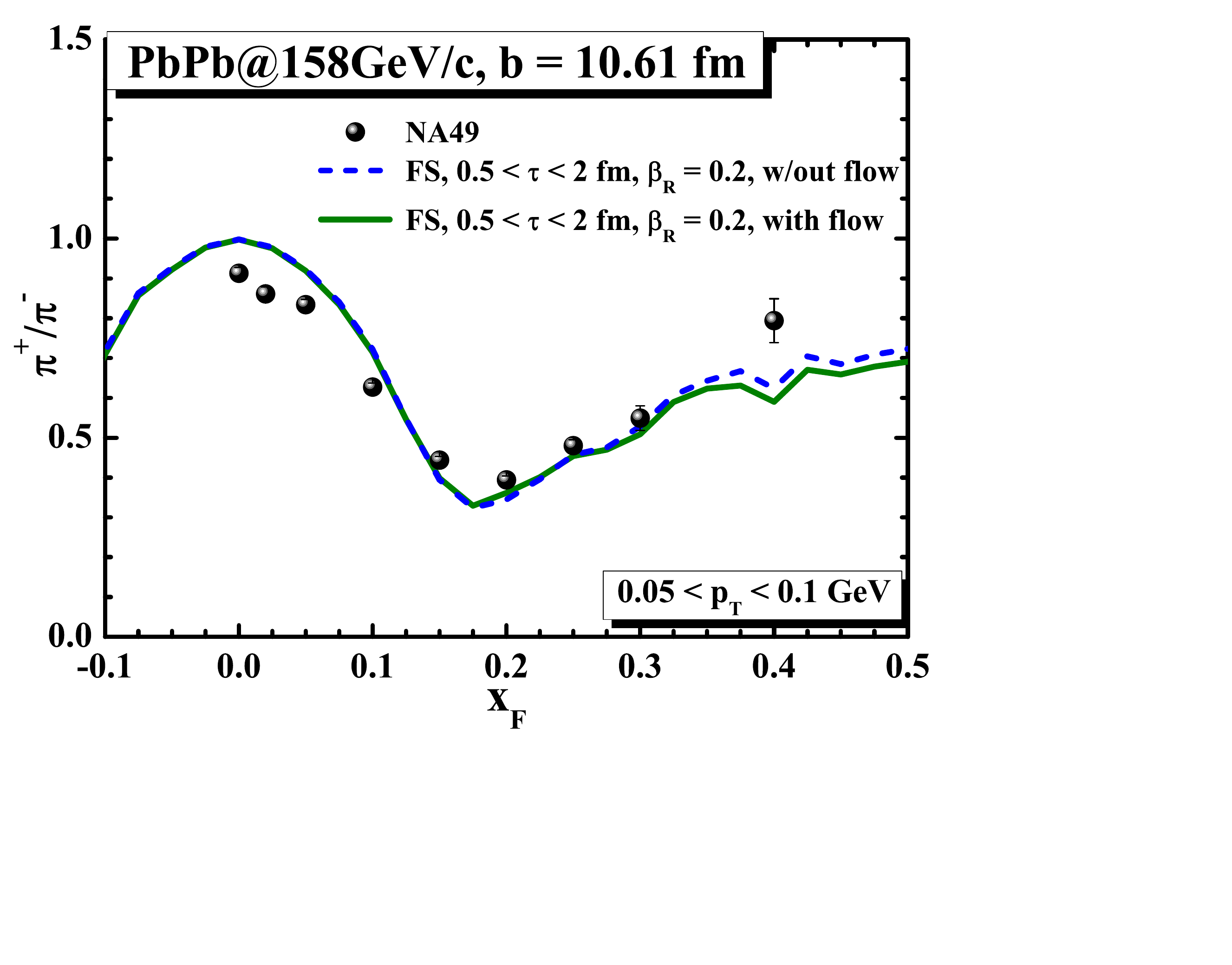}\hspace*{-6cm}
} } \vspace{-1.7cm}\caption{The effect of inclusion of the initial
$v_1$ and $v_2$ flow coefficients on the observed $\pi^+/\pi^-$
ratio (solid green lines). The dashed blue lines are for the case
when no flow is taken into account (the same as in
Fig.~\ref{final}).} \label{fig:ratio_flow}
\end{figure*}

\section{Results}
\vspace{-0.35cm}
First we perform the simulations for various fixed values of the pion creation time, $\tau$: $\tau=0.5; 1; 1.5; 2$~fm/$c$ (please note that this way we assume that the pion creation time is the same for each fire streak, in its rest frame). We figure out that there is no configuration with fixed pion creation time that can well describe the experimental data. In fact we expect that the fire-streak lifetime increases with its excitation energy. For this reason, we try to simulate an initial configuration with
the pion creation time which is not fixed. We choose the following simple linear dependence~\cite{EM_paper}:
\begin{equation}
\tau=a(E_s^*-m_s)+\tau_0, \label{tau_dependence}
\end{equation}
where $\tau_0=\tau_{min}=0.5$~fm/$c$ and $\tau_{max}$ is set to be $2$~fm$/c$, which gives us $a\approx0.08$.

In Fig.~\ref{final} by the solid red lines we present the results of calculation of the electromagnetic effect on the $\pi^+/\pi^-$ ratio as a function of $x_F$ in peripheral $Pb+Pb$ collisions at top SPS energies, with the pion creation time parametrized as in Eq.~(\ref{tau_dependence}). The results are shown for two different ranges of pion transverse momentum: $0<p_T<0.05$~GeV/$c$ (left plot) and $0.05<p_T<0.1$~GeV/$c$ (right plot). At this point we notice that in the calculations made so far we always assumed that spectators are {\em stable}, at least on the time scale when electromagnetic fields interact with charged pions (in our simulation the spectator systems were taken as two stable, homogeneously charged spheres as described in Ref.~\cite{EM_previous_1}). However, the spectators are rather
highly excited systems. Therefore, presently we impose a scenario with expansion of the spectators. In its own rest frame, each of the two spectator systems is taken as a homogeneously charged sphere, expanding radially with a given surface velocity $\beta_R$. We introduce this surface velocity $\beta_R$ as an additional free parameter. The configuration with $\tau$ given by Eq.~(\ref{tau_dependence}) and $\beta_R=0.2$ (in the spectator rest frame) gives the best description of the NA49 experimental data~\cite{NA49_EM} for both ranges of pion transverse momentum, $0<p_T<0.05$~GeV/$c$ and $0.05<p_T<0.1$~GeV/$c$, as shown in Fig.~\ref{final} by the dashed blue lines.

In Fig.~\ref{fig:ratio_flow} we demonstrate the effect of the pion flow on the observed $\pi^+ / \pi^-$ ratio as a function of Feynman $x_F$, for $v_1$ and $v_2$ included together. We find no effect for $v_2$, but a non-negligible effect for $v_1$.

\begin{figure*}
\vspace{0.3cm} \centering \subfigure{
\resizebox{0.47\textwidth}{!}{%
 \includegraphics{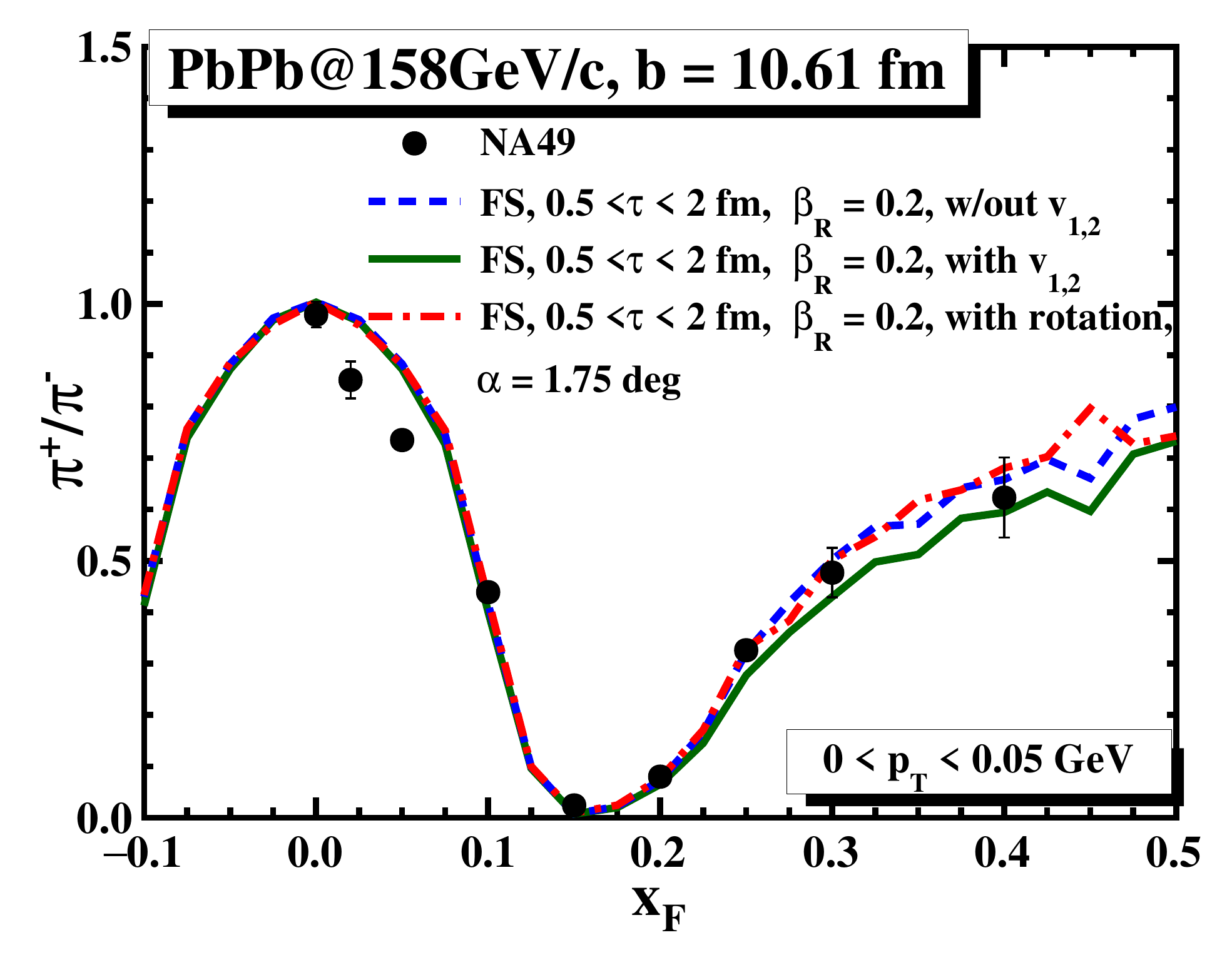}\hspace*{-1.3cm}
} } \subfigure{
\resizebox{0.47\textwidth}{!}{%
 \includegraphics{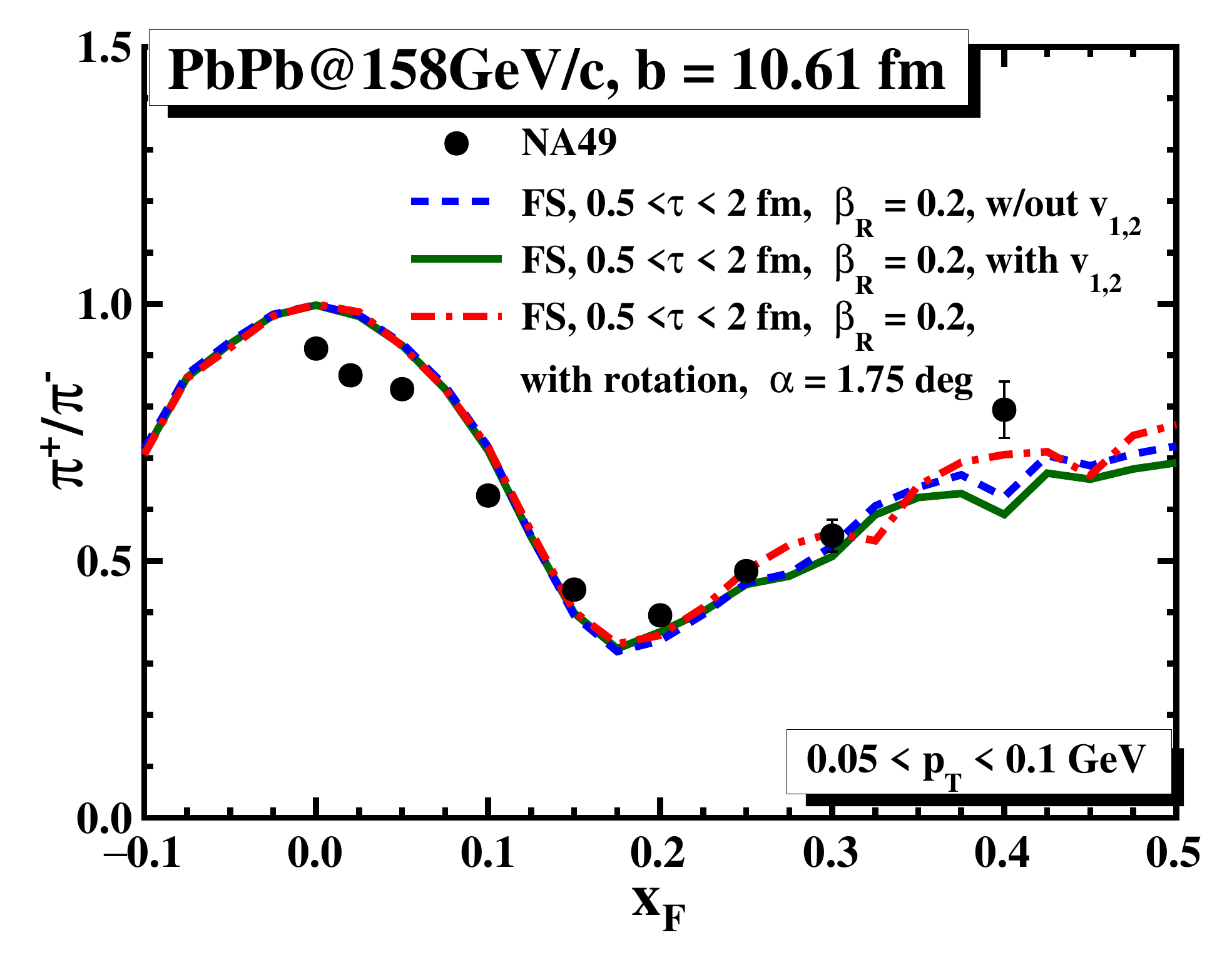}\hspace*{-1.3cm}
} } \vspace{-0.2cm}\caption{The effect of inclusion of fire-streak
rotation as described in the text on the observed $\pi^+/\pi^-$ ratio
(dash-dotted red lines). The dashed blue lines are for the case when no flow is taken into account and the green solid lines are for the case when flow is taken into account (both dashed blue and green solid lines are the same as in Fig.~\ref{fig:ratio_flow}).}
\label{fig:ratio_rotation}
\end{figure*}

Another effect to be considered is the well-known phenomenon of vorticity of strongly-interacting matter created in the collision~\cite{star-nature}. We estimate this effect in the following way. We assume that starting from the moment of closest approach of two nuclei ($t=0$) the fire streak rotates for a given small angle $\alpha$. After the rotation the fire streak follow its modified trajectory until pions are emitted from the fire streak. Consequently the pion emission point is shifted in transverse and longitudinal direction with respect to the case with no rotation. The size of the shift increases with increasing $\tau$ and $\alpha$. We find that rotation by $\alpha=1.75\deg$ gives a good description of the experimental data on the directed flow~\cite{EM_paper}. Finally, in Fig.~\ref{fig:ratio_rotation} (dash-dotted curve) we present the result of simulation made for the same values of $\tau$ (Fig.~\ref{fig:ratio_flow}, blue dashed curve) but including the rotation of fire streaks by a total angle $\alpha=1.75\deg$. Evidently no significant change is visible with respect to the case with no rotation also shown in the figure (dashed curve). This
is due to the small value of the angle $\alpha$ allowed for our model by the experimental data on $v_1$, which also results in a small displacement of the pion emission points. We conclude that for the relatively short pion emission times $\tau$ discussed here, vorticity has no much effect on the electromagnetic distortion of $\pi^+/\pi^-$ ratios.

\section{Summary}
\vspace{-0.35cm}
In the present contribution we introduced the model of the longitudinal evolution of the system, which explains the centrality dependence of pion yields and rapidity spectra in $Pb+Pb$ collisions~\cite{FS_model_1,FS_model_2} and provides the first ever quantitative description of the electromagnetic effects on $\pi^+/\pi^-$ ratio measured at large rapidity. 
We investigated that rather small pion creation times have been necessary to describe the experimental data on electromagnetic effects on fast pions ($0.5<\tau<2$~fm/$c$). The configuration with the expanding spectators gives the best description of the data. The inclusion of directed flow gives a non-negligible effect, whereas elliptic flow shows no effect. We also found that rotation (vorticity) of fire streaks
results in the presence of rapidity-dependent directed flow, but brings no effect on the spectator-induced electromagnetic modification of $\pi^+/\pi^-$ ratios.

\acknowledgments
\vspace{-0.35cm}
This work was supported by the National Science Centre, Poland under Grant No. 2014/14/E/ST2/00018.


\end{document}